\documentclass[usenatbib,usegraphicx,letterpaper]{mn2e}
\usepackage[totalwidth=480pt,totalheight=680pt]{geometry}

\usepackage{amssymb}
\usepackage{epsfig}
\usepackage{amsmath}

\newcommand{\beq}{\begin{equation}}
\newcommand{\eeq}{\end{equation}}
\newcommand{\ben}{\begin{enumerate}}
\newcommand{\een}{\end{enumerate}}
\newcommand{\bit}{\begin{itemize}}
\newcommand{\eit}{\end{itemize}}
\newcommand{\beqray}{\begin{eqnarray}}
\newcommand{\eeqray}{\end{eqnarray}}

\newcommand{\mr}{M_{\mathrm{r}}}

\newcommand{\monetwo}{m_{12}}
\newcommand{\bcg}{\mathrm{BCG}}

\newcommand{\sigmavred}{\sigma_v^{\mathrm{red}}}
\newcommand{\resid}{\delta\,\ln M}

\newcommand{\vmax}{V_{\mathrm{max}}}
\newcommand{\vacc}{V_{\mathrm{max}}^{\mathrm{acc}}}

\newcommand{\vl}{V_{\mathrm{L}}}
\newcommand{\ngal}{n_{g}}
\newcommand{\msun}{M_\odot}
\newcommand{\nh}{n_{h}}
\newcommand{\mhost}{M}
\newcommand{\lnmhost}{\ln M}
\newcommand{\mfit}{M_{\mathrm{fit}}}

\newcommand{\Omegam}{\Omega_{M}}

\newcommand{\littleh}{h}

\newcommand{\tilt}{n_{s}}
\newcommand{\sigmaeight}{\sigma_{8}}

\begin{document}

\title[Mind the Gap]
{Mind the Gap: Tightening the Mass-Richness Relation with Magnitude Gaps}

\author[A.P. Hearin et al.]
{Andrew P. Hearin$^1$,
Andrew R. Zentner$^1$,
Jeffrey A. Newman$^1$,
Andreas A. Berlind$^2$ \\
$^1$ Department of Physics and Astronomy \& \\
Pittsburgh Particle physics, Astrophysics and Cosmology Center (PITT PACC),\\ 
University of Pittsburgh, Pittsburgh, PA 15260;\\
 aph15@pitt.edu, zentner@pitt.edu, janewman@pitt.edu\\
$^2$ Department of Physics and Astronomy, Vanderbilt University, Nashville, TN;
a.berlind@vanderbilt.edu
}

\maketitle

\begin{abstract}

We investigate the potential to improve optical tracers of 
cluster mass by exploiting measurements of the magnitude gap, 
$\monetwo,$ defined as the difference between the r-band absolute magnitude 
of the two brightest cluster members. We find that in a mock sample of galaxy 
groups and clusters constructed from the Bolshoi simulation, the scatter about 
the mass-richness relation decreases by $\sim 15-20\%$ when magnitude 
gap information is included. A similar trend is evident in a volume-limited, spectroscopic 
sample of galaxy groups observed in the Sloan Digital Sky Survey (SDSS). We find that 
SDSS groups with small magnitude gaps are richer than large-gap groups at fixed values 
of the one-dimensional velocity dispersion among group members $\sigma_v,$ which we 
use as a mass proxy. We demonstrate explicitly that $\monetwo$ contains information 
about cluster mass that supplements the information provided by group richness and 
the luminosity of the brightest cluster galaxy, $L_{\mathrm{BCG}}$.  In so doing, we 
show that the luminosities of the members of a group with richness $N$ 
are inconsistent with the distribution of luminosities that results from $N$ random 
draws from the {\em global} galaxy luminosity function.  As the cosmological 
constraining power of galaxy clusters is limited by the precision in cluster 
mass determination, our findings suggest a new way to improve 
the cosmological constraints derived from galaxy clusters.

\end{abstract}


\section{Introduction}
\label{section:intro}

Galaxy clusters have long been exploited to probe the composition of the universe.  
The utility of galaxy clusters as cosmological probes using a broad range 
of techniques has been reviewed extensively by \citet{allen_etal11_review}.  
Galaxy cluster observations are a key component of any 
effort to constrain the cause of cosmological acceleration \citep[see the review by][]{weinberg_etal12_review}.  
Among many advances in cluster cosmology, 
modern optical surveys have enabled the construction of large samples of optically-identified 
clusters which, in turn, have led to competitive cosmological constraints from optically-identified 
cluster abundances \citep[e.g.,][]{gladders_etal07,rozo_etal10}.  

Much of the constraining power of clusters results from determinations of their abundance 
as a function of their mass.  The abundance of clusters by mass may be 
reliably predicted \citep[e.g.,][]{tinker_etal08}, but cluster mass is not directly observable.  
Optical cluster cosmology efforts generally rely on using the number of cluster members ({\em richness}) as a 
proxy for mass to concurrently fit for cosmological parameters and the mass-richness relation  
\citep[although, see][for another technique]{newman_etal02}.  
In detail, richness must be defined precisely for the observational sample under consideration, so that the   
specific definitions of richness vary depending upon survey characteristics and cluster identification methods.  

One way to improve cosmological constraints from optically-identified clusters is to reduce the 
scatter between the observable (richness) and the predicted quantity (mass) \citep{rozo_etal10,rykoff_etal12}.  In this paper, we suggest that the differences in absolute magnitude between 
the most luminous cluster members, data already 
contained within optical surveys aiming to perform cluster cosmology, 
can be harnessed to reduce the scatter in cluster mass for a fixed set of observables. 
Specifically, we show that {\em magnitude gap}, the difference in r-band absolute magnitude between the 
brightest and second brightest members of a galaxy group, can aid in the determination 
of group mass at both fixed richness and fixed r-band luminosity of the brightest group member.  
We provide theoretical and observational support for this suggestion using 
the Bolshoi simulation of cosmological structure growth \citep{klypin_etal11} and galaxy group and 
cluster data from Data Release 7 of the Sloan Digital Sky Survey (SDSS) \citep{sdss_dr7}.  
In \S~\ref{section:sims} we describe mock galaxy catalogs constructed from the 
Bolshoi simulation. A brief description of the SDSS group data is given in \S~\ref{section:data}. 
We present results from our mock galaxy catalog in \S~\ref{section:predictions} 
and from our analysis of the SDSS groups in \S~\ref{section:observations}.  
We draw brief conclusions from these results in \S~\ref{section:conclusions}.

\section{Simulations \& Mock Galaxy Catalogs}
\label{section:sims}

We use the Bolshoi N-body simulation \citep{klypin_etal11} to study 
the connection between magnitude gaps within galaxy groups and 
the mass-richness relation.  Bolshoi models the growth of structure 
in a cubic volume $250\,\littleh^{-1}\mathrm{Mpc}$ on a side within a 
$\Lambda$CDM cosmology with total matter density $\Omegam=0.27$, Hubble constant 
$\littleh=0.7$, power spectrum tilt $\tilt=0.95$, and power spectrum normalization $\sigmaeight=0.82.$  
The Bolshoi data are available at {\tt http://www.multidark.org} and we refer the 
reader to \citet{riebe_etal11} for additional information.
Our analysis requires reliable identification of self-bound 
subhalos within virial radii of distinct halos. We utilize the 
{\tt ROCKSTAR} \citep{behroozi_etal11} halo finder in order to identify 
halos and subhalos within Bolshoi.

To connect the properties of galaxies to dark matter halos we employ 
the widely used subhalo abundance matching technique (SHAM) 
\cite[e.g.,][]{kravtsov_etal04,conroy_etal06}. 
We assume a monotonic relationship between 
the r-band luminosities of galaxies and the maximum circular speeds 
of test particles within their host dark matter halos, 
$\vmax\equiv\mathrm{max}\left[\sqrt{GM(<r)/r}\right]$, where $M(<r)$ is the mass of the halo interior to the radial coordinate $r.$ 
In broad terms, the reasoning behind this algorithm is that r-band luminosity 
is a rough proxy for stellar mass and stellar mass should correlate with the 
depth of the gravitational potential well, described by $\vmax$.  
Subhalos evolve significantly due to interactions upon incorporation into a larger 
distinct halo, so $\vmax$ for subhalos may be a poor proxy for stellar mass or r-band 
luminosity.  \citet{conroy_etal06} demonstrated that a SHAM algorithm 
that assigns luminosities to subhalos based upon the maximum circular speed of 
the subhalo {\em at the time it merged with the distinct halo}, $\vacc$, can 
describe a broad range of galaxy clustering data from $z\approx 0$ to $z\approx 4$.  
We define the circular speed used in our luminosity 
assignment to $\vl = \vmax$ for distinct halos and $\vl = \vacc$ for subhalos.  
We assign r-band luminosities to halos and subhalos through the implicit relation 
\beq
\label{eq:lv}
\ngal(>L)=\nh(>\vl), 
\eeq
where $\ngal(>L)$ is the number density of observed galaxies with r-band 
luminosity $>L$  \citep{blanton_etal05} and $\nh(>\vl)$ is the number density of dark matter halos 
and subhalos with circular speed $>\vl$.  Eq.~(\ref{eq:lv}) ensures that the 
distribution of luminosities assigned to dark matter halos and subhalos 
matches the observed luminosity function of galaxies. SHAM models of this 
kind successfully describe a variety of astronomical 
data \citep[see][and references therein]{klypin_etal11,trujillo-gomez_etal11,watson_etal11}.

Once brightnesses have been assigned to halos according to Eq.~(\ref{eq:lv}), 
we construct mock galaxy samples by imposing a brightness cut $M_r<-18$ on all the mock galaxies  
in Bolshoi. This brightness cut corresponds to $\vmax> 92$ km/s, well above the $50$ km/s completeness limit of Bolshoi \citep{klypin_etal11}.
 In this mock catalog, we consider galaxy groups to be collections of subhalos 
associated with the same host halo. 
We study the properties of these mock groups in \S~\ref{section:predictions}. 

\section{Observational Data}
\label{section:data}

To study the mass-richness relation observed in low-redshift groups and clusters 
we use a volume-limited catalog of galaxy groups identified in Data Release 7 of 
the SDSS using the algorithm described in \citet{berlind_etal06}. This is an update of the \citet{berlind_etal06} groups 
(based on SDSS Data Release 3) to SDSS Data Release 7.  
All of the members of this sample are members of the main 
galaxy sample of SDSS Data Release 7.  The group catalog is constructed using a redshift-space 
friends-of-friends algorithm that has been corrected for incompleteness due to fiber collisions. 
The particular group catalog we use is constructed from galaxies in a volume-limited spectroscopic 
sample in the redshift range $0.02 < z < 0.068$ with r-band absolute magnitude 
$M_r - 5\log h < -19$. We refer to this catalog as the ``Mr19'' group catalog. 
Each of the $6439$ groups in the Mr19 catalog contains $N>2$ members.  
We refer the reader to \citet{berlind_etal06} for further details on the 
group finding algorithm. 

\section{Predictions for the Potential Utility of Gap Information}
\label{section:predictions}

In this section we use the mock galaxy catalog described in \S~\ref{section:sims} 
to study the predictions of abundance matching for the dependence of host halo mass on 
galaxy group richness and magnitude gap.  Such a study is idealized for several reasons, including  
 the relative simplicity of the SHAM algorithm and the fact that groups can be unambiguously  
identified with halos of a particular mass.  Nevertheless, this demonstration has the distinct 
advantage that the masses of the host halos in the Bolshoi simulation are known and do not need 
to be inferred imperfectly from observational data.  To proceed, we define richness $N$ as the number 
of mock galaxies within the host halo brighter than our 
specified absolute magnitude threshold.  We take the magnitude gap to be the difference between the r-band 
absolute magnitude of the brightest galaxy within the halo and the second brightest galaxy 
within the halo, $\monetwo = M_{r,1} - M_{r,2}$, where $M_{r,i}$ is the r-band absolute magnitude of 
the $i^{\mathrm{th}}$ brightest galaxy in the halo.  

\begin{figure}
\centering
\includegraphics[width=9.0cm]{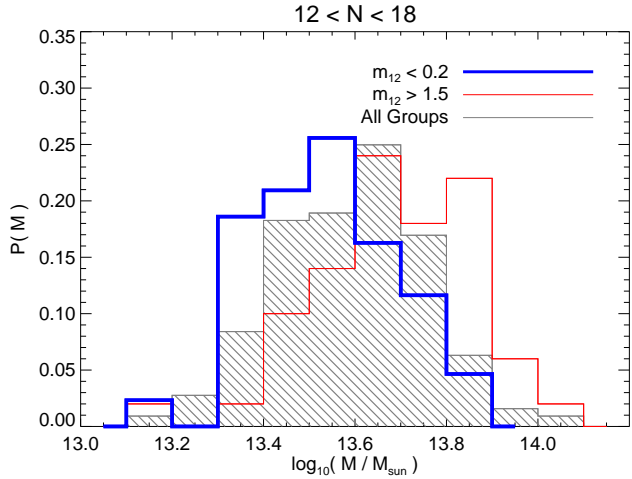}
\caption{
Plot of the mass distribution of host halos in Bolshoi in the richness range $12<N<18.$ The mass distribution for 
all host halos in this richness range appears as the hatched, gray histogram. The thin red curve traces this distribution for the subsample of host systems with a very large magnitude gap ($\monetwo>1.5$). The mass distribution of host halos with 
$\monetwo<0.2$ is plotted with the thick blue curve. Note that most objects are not in either $\monetwo$ bin traced by the red or blue histograms, which is why the full distribution plotted in gray does not resemble a combination of the red and blue. Evidently, large gap systems tend to be more massive at fixed richness, 
suggesting the possibility that $\monetwo$ can be used to improve the calibration of the mass-richness relation.
}
\label{fig:massdist}
\end{figure}

We illustrate our motivation for exploring the utility of gap information in Figure~\ref{fig:massdist}, 
in which we plot the mass distributions of host halos in Bolshoi in a narrow range of richness, 
$12<N<18$.  The gray, hatched histogram traces the mass distribution for {\em all} host halos 
in this richness range, while the thin red (thick blue) curve pertains to host systems with a large 
(small) magnitude gap $\monetwo.$ Evidently, large gap systems tend to be more massive than their small gap counterparts 
at fixed richness. Of course the trend in Fig.~\ref{fig:massdist} could simply be due to the finite width of the richness bin we have chosen, 
rather than following as a consequence of the mass-richness residual being correlated with $\monetwo.$ 
To explore this issue more rigorously, we employ standard regression analysis 
techniques to find the linear relationship between $\mathrm{ln}(N)$ and $\mathrm{ln}(\mhost)$ 
that minimizes $\sigma(\lnmhost).$ The use of a linear regression is well-motivated by 
previous results \citep[for example,][]{becker_etal07} that find the mass-richness relation to be well-described 
by a power law.  We find that our best fit model, $\mfit(N)= CN^{\alpha}$, 
with $C=2.2\times10^{12}M_{\odot}$, and $\alpha=1.1$, 
gives an accurate description of the mass-richness relation for the rich groups ($N\ge10$) in our mock sample, 
yielding a mean residual $\langle \delta \ln \mhost \rangle \simeq0.02$, 
and a residual dispersion of $\sigma(\lnmhost) \simeq 0.33$.

\begin{figure}
\centering
\includegraphics[width=9.0cm]{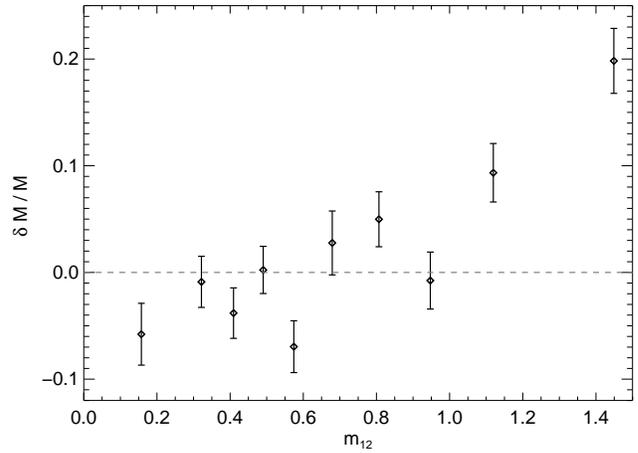}
\caption{
Plot of the residual $\resid$ of the best-fit mass-richness power law relation exhibited by 
groups and clusters in Bolshoi as a function of $\monetwo.$ A linear fit 
to the residuals indicates that $\resid \propto 0.18\,\monetwo$, suggesting 
that exploiting the magnitude gap may significantly improve cluster mass estimation 
techniques that rely only on richness. 
}
\label{fig:residual}
\end{figure}

In Figure~\ref{fig:residual} we plot the mean residual $\resid$ as a function of $\monetwo.$  
The trend suggested by Fig.~\ref{fig:massdist} is borne out: groups and clusters with a 
large (small) magnitude gap $\monetwo$ are more (less) massive than the average 
$\mhost$ at a given richness. A linear fit to the results illustrated in Fig.~\ref{fig:residual} 
indicates that $\resid \propto 0.18\,\monetwo$, 
implying that there may be significant information about a cluster's 
mass contained in the magnitude gap $\monetwo$ that is not contained in the richness alone. 
Moreover, the mass-dispersion about the $\monetwo-N$ plane determined by a two-dimensional linear regression improves by $18\%$ to $\sigma(\lnmhost) \simeq 0.27,$ further demonstrating the potential improvement in mass determination provided by the use of magnitude gap information.

While the trend in the mass-richness relation with $\monetwo$ as 
seen in Fig.~\ref{fig:residual} is the novel feature of this work, 
the tendency for large gap systems to be more massive at fixed richness 
appears in a variety of guises in the literature on {\em fossil groups}, 
which are typically defined to be galaxy groups with $\monetwo > 2$.  
The picture of fossil groups that is the most prevalent in the literature 
is that these are groups which assembled most of their mass at high redshift, 
so that processes such as dynamical friction and mass loss have had ample 
time to deplete these systems of their most massive satellite galaxies, 
leaving behind a very bright central galaxy with few comparably bright satellites 
\citep{jones_etal03,donghia_etal05,zentner_etal05}\footnote{See also \citet{skibba_etal11}, who studied a closely related issue: the relative brightness of a group's most luminous satellite and central galaxy.}.

In the above scenario for the origins of fossil groups, 
the same processes that lead to the formation of a large magnitude gap are also at work 
in the depletion of the number of group members above a given brightness threshold. 
In other words, this picture posits a dynamical connection between $\monetwo$ and richness.  
However, even in the complete absence of such dynamical processes, we may still expect 
large-gap systems to have fewer members than small-gap systems. This is a 
consequence of the shape of the Schechter function: for any luminosity function $\Phi(L)$ 
with a slope that steepens with brightness, the average gap $\monetwo$ obtained from a set 
of $N$ random draws from $\Phi(L)$ increases as $N$ decreases. 

The extent to which dynamical processes influence the relationship between $\monetwo$ 
and $N$ remains an open question, but we point out here that if $\monetwo$ were purely 
statistical, resulting from $N$ random draws from a Schechter function, the magnitude gap 
would contain no information about mass that would not already be provided by knowledge of 
richness. This follows from Bayes' Theorem. If $\monetwo$ were determined strictly by $N$ 
random draws from a common luminosity function, then $P(\monetwo|M,N)=P(\monetwo|N)$. Then 
it follows directly from Bayes' Theorem that $P(M|N,\monetwo)=P(M|N)$; that is, the probability distribution of 
mass is unchanged by knowledge of $\monetwo$ when the richness is known. 
This is sensible because in this scenario the $\monetwo$ distribution is 
determined entirely by $N$, so knowledge of $\monetwo$ provides no new, 
independent information about the system. Thus the trend exhibited by the 
mass-richness relation seen in Fig.~\ref{fig:residual} reflects a relationship 
between the magnitude gap and the mass of host halos in our mock sample beyond 
that expected from sampling a luminosity function a finite number of times.

The SHAM-based results presented in this section show that a simple 
and well-supported model for populating halos with galaxies makes a definite 
prediction for the relationship between halo mass, magnitude gap, and richness in groups.  
To be sure, there are many factors that will tend to wash out the clear trend seen in 
Fig.~\ref{fig:residual}. For example, group/cluster membership (and hence, richness) is 
determined by halo membership, so contamination by interlopers due 
to projection effects is not included; the simulations do not suffer 
from edge effects that may alter the richnesses of groups in real surveys; and 
the fidelity with which the SHAM prescription for identifying 
halos and subhalos with luminous galaxies correctly describes richness 
has not been extensively tested in this context.  As we will see in the 
next section, despite these complicating factors the dependence of the mass-richness 
relation on $\monetwo$ appears to be significant in a spectroscopic sample of 
galaxy groups observed in SDSS.

\section{The Observed Connection between Mass, Richness, and Magnitude Gap}
\label{section:observations}

We now examine the mass-richness scaling relation seen in groups and 
clusters in the volume-limited Mr19 galaxy group catalog described 
in \S~\ref{section:data}. The principal result of this section appears in the bottom right panel of Figure ~\ref{fig:upshot}, which we will argue demonstrates that the magnitude gap contains information about cluster mass that is independent from both richness and $L_{\bcg}.$ Most of the work described in this section, illustrated in the remaining three panels of Fig.~\ref{fig:upshot}, addresses a variety of possible systematics and selection effects that are germane to our primary conclusion.

As our mass proxy for the groups we use $\sigma_v$, 
the one-dimensional velocity dispersion of member galaxies, defined as:
\beq
\sigma_v \equiv \frac{c}{1+\bar{z}}\sqrt{\frac{1}{N-1} \sum_{i=1}^{N} (z_{i}-\bar{z})^{2} },
\eeq
where $N$ is the number of group members, $z_i$ are the redshifts of the 
member galaxies, and $\bar{z}$ is the redshift of the (unweighted) group centroid.\footnote{Note that in \S~\ref{sub:sigmavnoise} we show that our results are unchanged by whether or not one assumes that the central group galaxy is at rest in the calculation of $\sigma_v.$}
We are chiefly interested in determining whether the scaling of richness with $\sigma_v$ 
changes when comparing samples of groups with different $\monetwo$, 
as is suggested by the results in \S~\ref{section:predictions}.  
Assigning $\monetwo$ values to the groups requires some care due to complications 
presented by fiber collisions which affect $\sim 8\%$ of the galaxies in Mr19.  
Fiber collided galaxies in the Mr19 sample are assigned the redshift of their 
nearest neighbor on the sky. As a galaxy's redshift is used to infer its absolute 
magnitude, this can result in a large error in the inferred brightnesses of 
fiber collided galaxies. Thus if either the brightest member of the group (BCG) 
or the second-brightest member (which we denote below as the SBG for `second-brightest galaxy') 
is fiber collided, the group will be assigned 
an erroneous value of $\monetwo$. As the BCG in particular is likely to be found 
near the group centroid where the galaxy number density can be very large, this scenario 
is relatively common; we find that $\sim 2\%$ of the groups in Mr19 have either a BCG or 
a SBG that is affected by fiber collisions.  We define $\monetwo$ to be the magnitude difference 
between the two brightest non-fiber-collided galaxies in the group, but we note 
that all of our conclusions remain unchanged if we allow galaxies affected by fiber collisions 
to be used in the definition of the magnitude gap.

In the upper left panel of Figure \ref{fig:upshot} we plot the mean 
richness of galaxy groups in Mr19 as a function of $\sigma_v$.  This panel illustrates 
the results of a calculation in which we have divided the Mr19 groups into 
logarithmically-spaced bins over the range $150\, \mathrm{km/s} \le \sigma_v \le 350\, \mathrm{km/s}$, 
and computed the mean richness in each bin.  Evidently, large-gap systems do indeed appear to 
be less rich at fixed $\sigma_v$ (mass) than small-gap systems.  However, prior to 
drawing this conclusion we now address several possible selection effects and complicating 
factors that could influence this result.  The primary result of this section 
is shown in the lower, right panel of Fig.~\ref{fig:upshot} in which we exhibit the 
$\sigma_v-N$ relation after controlling for the uninformative, intrinsic correlations 
between $\monetwo$ and richness as well as $\monetwo$ and the absolute magnitude of the 
brightest galaxy in the group.

Before proceeding, we draw attention to a contrast between our presentation of the 
predictions of the standard cosmological model in \S~\ref{section:predictions}, and 
our present analysis of the observed SDSS DR7 groups.  In the present analysis, we begin by 
presenting results by binning the sample according to the mass proxy ($\sigma_v$) rather 
than richness, which we have found gives a clearer demonstration of our principal result. 
However, this choice introduces several possible systematics (for example, errors on $\sigma_v$ 
are correlated with richness). In \S~\ref{sub:sigmavnoise} we address these systematics in two distinct ways. 
First, we verify that the conclusions we draw based on our principal methodology are robust to systematic errors pertaining to noisy measurements of $\sigma_v.$ Second, at the end of  \S~\ref{sub:sigmavnoise} we provide an alternative demonstration of our conclusions by instead binning our sample according to the richness $N.$ Regardless of whether we bin on richness or our mass proxy, our conclusions remain the same: large-gap groups exhibit a different mass-richness scaling relation than small-gap groups, with large-gap groups being more massive at fixed richness than small-gap groups.

\subsection{Controlling for the Magnitude Gap-BCG Correlation}

Selecting groups with a large magnitude gap biases one to select 
groups with a luminous BCG. For example, the Mr19 galaxy sample 
contains no galaxies dimmer than $M_r=-19,$ so selecting groups with 
$\monetwo>1.5$ requires the BCG to have a brightness $M_r<-20.5$.  
Meanwhile, \citet{reyes_etal08} suggested that BCG luminosity 
provides information on group mass that is independent of richness, so 
we must account for this bias to ensure that the magnitude gap provides 
information that is independent from the known correlation between 
mass and BCG luminosity.

We illustrate the differences in BCG luminosity induced by this selection 
in the lower left panel of Fig.~\ref{fig:upshot}, with the thin blue (thick red) 
histogram tracing $\Phi_{\bcg}(L)$ for small-gap (large-gap) systems.  The upper right panel of 
Fig.~\ref{fig:upshot} is an illustration of the potential importance of 
this effect as BCG luminosity clearly informs the $\sigma_v-N$ relation 
of the SDSS galaxy groups.  Systems with brighter BCGs are also richer 
at fixed $\sigma_v$.  This trend has the sense that should be expected if 
$L_{\bcg}$ were determined by the brightest of $N$ random draws from a fixed, global 
luminosity function \citep[e.g.,][]{paranjape_sheth11}.  

To account for bias from differences in the BCG luminosity distributions between 
large-gap and small-gap groups, we have drawn a random 
subsample of $1000$ of the low-gap, $\monetwo<0.2$, 
groups with a BCG brightness distribution 
that matches that of the large-gap, $\monetwo>1.5$ groups.  
The magenta, hatched histogram in the bottom left panel of Fig.~\ref{fig:upshot} shows 
$\Phi_{\bcg}(L)$ of the resulting $L_{\bcg}$-matched subsample of the $\monetwo<0.2$ groups.  
The $\sigma_{v}-N$ scaling relation for the small-gap, 
matched-BCG systems is shown as the purple triangles in the 
upper left panel of Fig.~\ref{fig:upshot}.  Eliminating any differences between the BCG 
luminosity distributions of low-gap and high-gap systems
 {\em increases} the disparity between the $\sigma_v-N$ relations of 
low- and high-gap groups.  We conclude that 
$\monetwo$ informs the $\sigma_v-N$ relation in a manner 
that is independent from BCG luminosity alone.

\subsection{Controlling for the Statistical Magnitude Gap-Richness Correlation}

\begin{figure*}
\centering
\includegraphics[width=8.0cm]{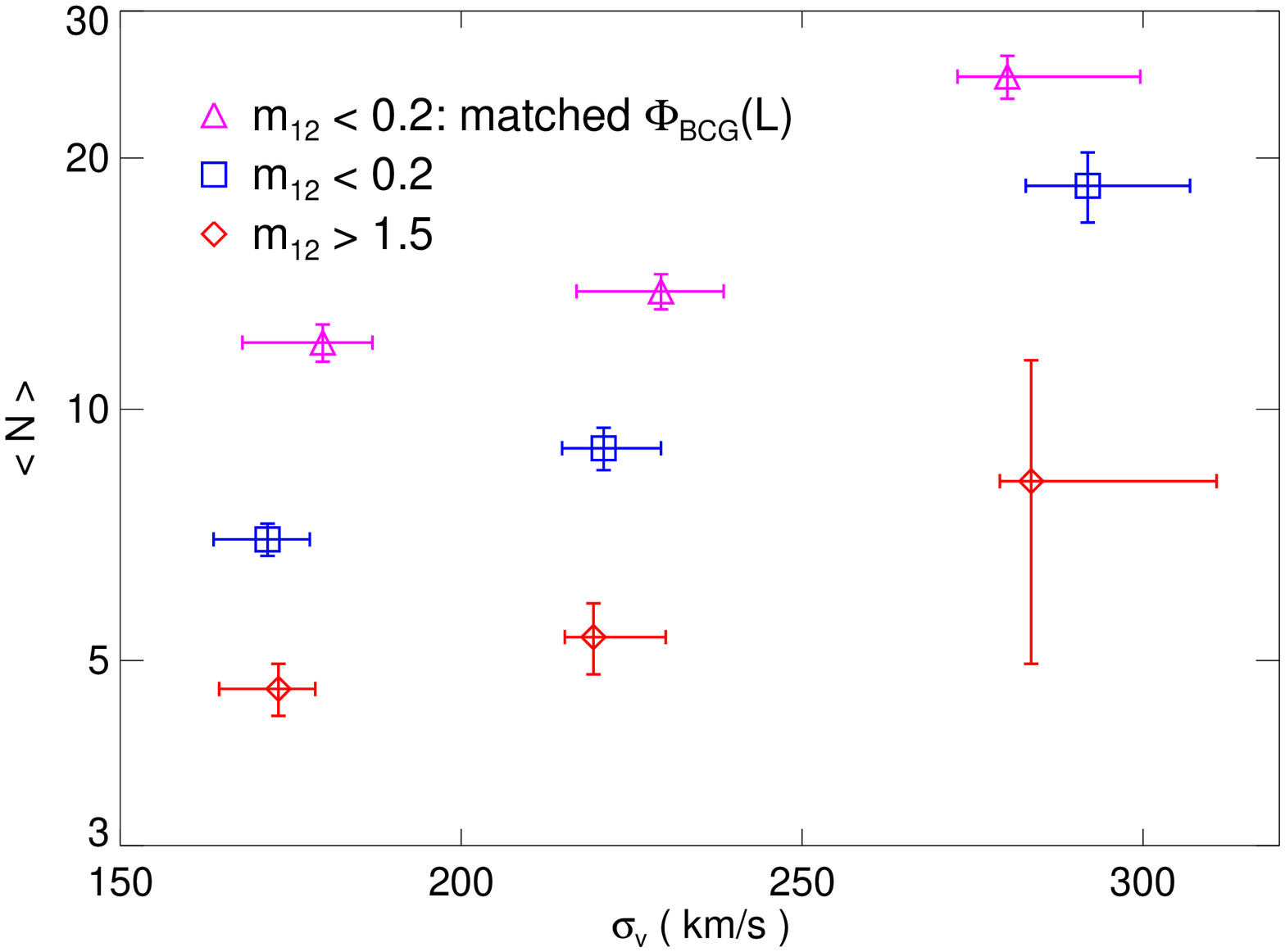}
\includegraphics[width=8.0cm]{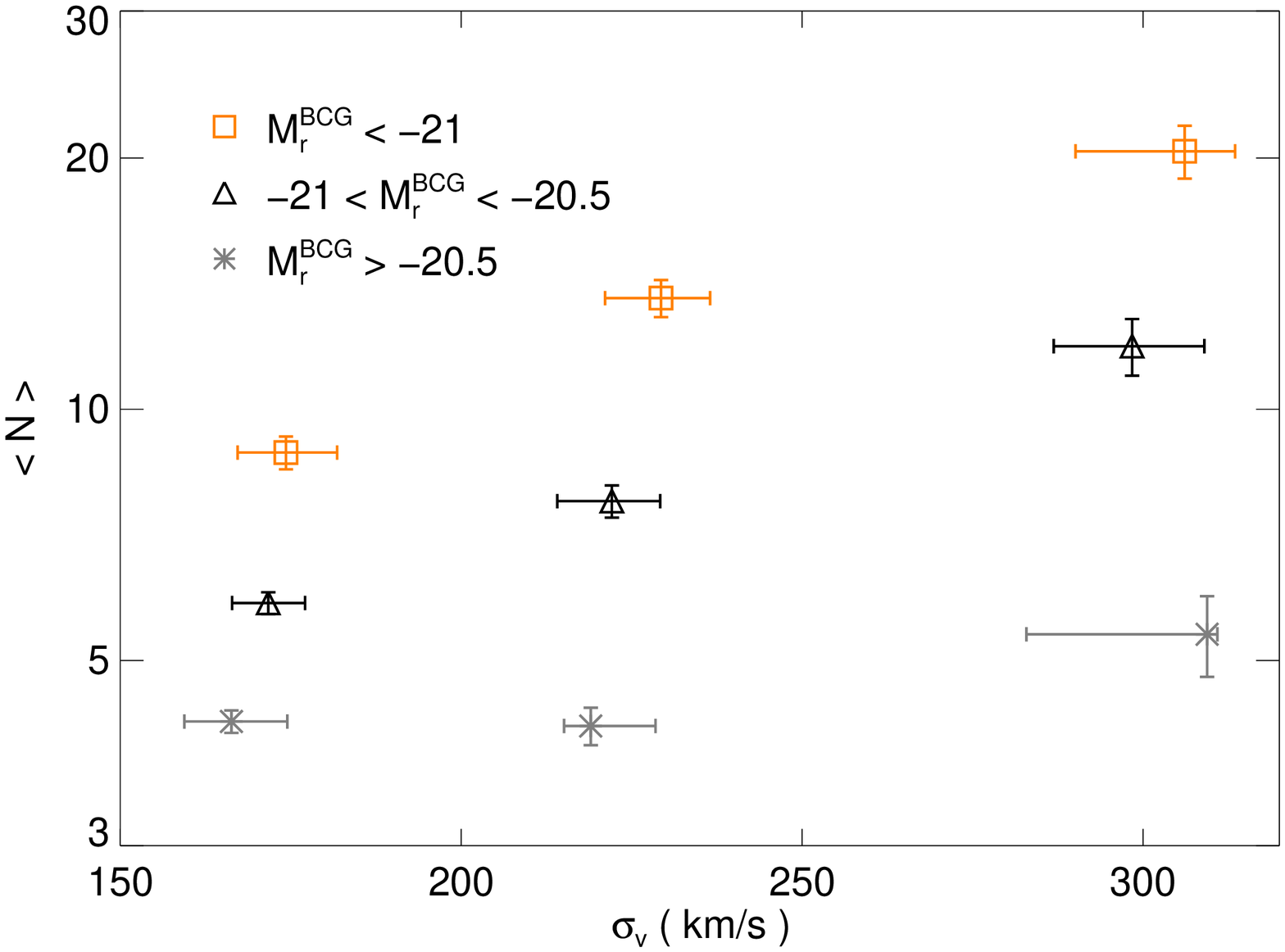}
\includegraphics[width=8.0cm]{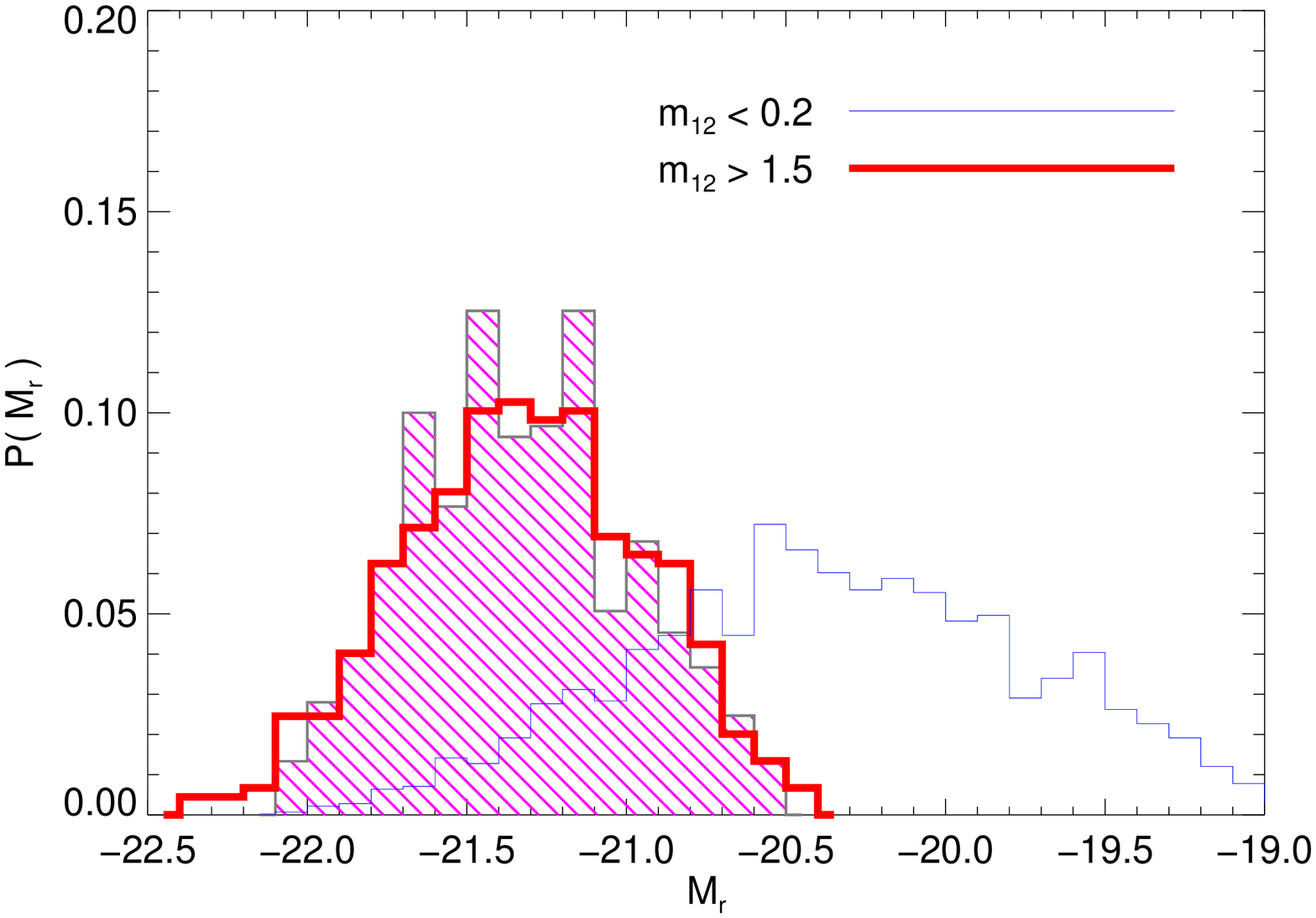}
\includegraphics[width=8.0cm]{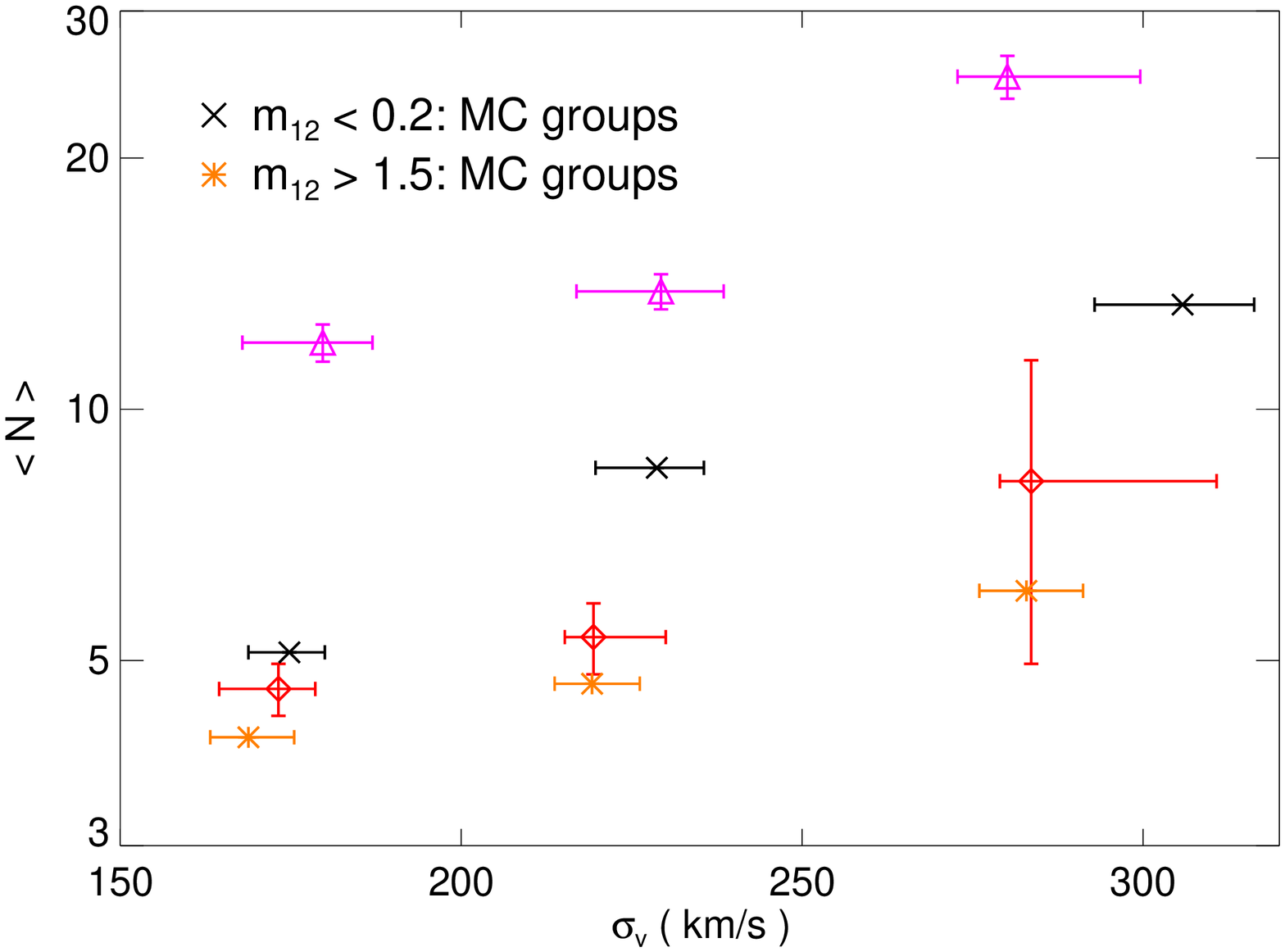}
\caption{
{\em Upper Left:} Richness as a function of mass proxy, $\sigma_v.$ 
Points are placed at the median $\sigma_v$ in each bin.  
Horizontal error bars represent the inner quartiles of the $\sigma_v$ distribution in each bin. Vertical 
error bars show the error on the mean. Red diamonds show systems 
exhibiting a large magnitude gap $\monetwo>1.5$, while blue squares show small-gap systems. 
The magenta triangles represent a randomly selected subsample of small-gap systems with 
a BCG luminosity function $\Phi_{\bcg}(L)$, matched to that of the large-gap systems.
{\em Lower Left:}  Thick (Thin) histograms show the distribution of 
BCG luminosity $\Phi_{\bcg}(L)$ (normalized to unit area) of large- (small-) gap 
systems in Mr19. The magenta, hatched histogram traces the 
BCG luminosity function of a random subsample of small-gap systems selected to match 
$\Phi_{\bcg}(L)$ of the large-gap systems. 
{\em Upper Right:} Mean richness as a function of $\sigma_v$ for galaxy groups with 
different BCG brightnesses.
{\em Lower Right:} The $\sigma_{v}-N$ scaling relations for large-gap groups and 
small-gap groups.  All samples have the same $\Phi_{\bcg}(L)$.  Orange asterisks and black crosses
represent results from Monte Carlo (MC) simulation of the group population in a model 
assuming that gap does not explicitly depend upon $\sigma_v$ (the group mass proxy).  
MC groups have negligible errors on the mean richness.
Red diamonds and magenta triangles represent large- and small-gap groups 
in the SDSS data, as in the upper left panel.  Differences between large-gap and small-gap groups in the data 
are significant compared to the MC groups, demonstrating that the magnitude 
gap $\monetwo$ contains information about group/cluster mass that is {\em independent} from 
{\em both} richness and $L_{\bcg}.$
}
\label{fig:upshot}
\end{figure*}

The principal result of our analysis in this section lies 
in the comparison between the red diamonds and magenta triangles in 
Fig~\ref{fig:upshot}: at fixed $L_{\bcg}$, large-gap systems are 
under-rich relative to small-gap systems at fixed $\sigma_v$ (mass). 
However, because of the natural correlation between 
$\monetwo$ and $N$ discussed in \S~\ref{section:predictions} in the 
context of our simulation analysis, some care is required 
before interpreting the differences between these points 
as implying that large-gap and small-gap systems have 
intrinsically different mass distributions 
at fixed richness. The difference between the red diamonds and magenta triangles 
shows that $P(N|\sigma_{v},\monetwo) \ne P(N|\sigma_{v})$, 
or equivalently, $P(\monetwo|\sigma_{v},N)\ne P(\monetwo|\sigma_{v})$.  
This inequality would hold even in a universe in which $\monetwo$ is 
determined solely by $N$ random draws from a global luminosity function. 
Yet, as discussed in \S~\ref{section:predictions}, in such a universe the 
magnitude gap contains {\em no} information about cluster mass that is {\em independent} 
from richness.  However, it is possible to show that the $\sigma_{v}-N$ scaling relations exhibited 
by the SDSS groups are distinct from the relations expected if $\monetwo$ were solely determined 
by random selection from a global luminosity function.  

We demonstrate that this is the case by contrasting the SDSS group data against the following 
Monte Carlo (MC) simulation. We draw a random galaxy group from our SDSS sample 
and assign the values of $\sigma_v$ and $N$ of this group to the ``MC'' galaxy group.
We populate the MC group by drawing $N$ galaxies from the global luminosity function of all the galaxies in the 
Mr19 group sample.  We assign the MC group a value of $\monetwo$ by taking the difference between 
the r-band absolute magnitudes of the brightest and next-brightest galaxies used to populate the 
MC group.  We repeat this procedure $10^6$ times to construct a sample of one million MC groups 
with the same group multiplicity function and $\sigma_{v}-N$ scaling relation as the Mr19 sample.   
The objective of this exercise is solely to construct mock samples of group 
galaxies with identical luminosity functions as the observed sample, but with 
magnitude gap determined solely by richness.

The MC groups exhibit a correlation between $\monetwo$ and $N$ at fixed $\sigma_v$, as expected. 
This is the correlation induced solely by the statistics of random draws from the group galaxy 
luminosity function. In the bottom right panel of Fig.~\ref{fig:upshot}, 
we plot the $\sigma_{v}-N$ scaling relation 
for MC groups with $\monetwo>1.5$ with orange asterisks, 
and MC groups with $\monetwo<0.2$ with black crosses.  
Both samples of MC groups have been selected so that their $L_{\bcg}$ distributions match 
that of the large-gap systems seen in the data, so that all the samples plotted in the 
bottom right panel have the same $\Phi_{\bcg}(L).$ 

Note that the scaling of richness with  $\sigma_v$ in the two samples MC groups 
is distinct even though $\monetwo$ cannot inform the velocity dispersion of the groups. 
This demonstrates the purpose of this exercise: if it is possible to use $\monetwo$ to 
inform mass-determination, then the trend in the mass-richness scaling relation seen 
in the data must be stronger than it is in the Monte Carlo sample, in which the gap is uninformative. 
Indeed, this is the case: the differences in richness at fixed $\sigma_{v}$ between the 
large-gap and small-gap groups in the SDSS data 
are clearly significant compared to their MC counterparts.
We conclude that the gap distribution is {\em not} determined solely by 
statistical effects, providing strong evidence that having a large magnitude gap is, 
in fact, correlated with being under-rich at a given mass.  Moreover, because we have 
controlled for the brightness of the BCG as described above, this implies that the magnitude gap $\monetwo$ 
contains information about the masses of galaxy groups that is {\em independent 
from both richness and BCG brightness}.  These results suggest that it may be 
possible to use $\monetwo$ to improve optical estimators of group and cluster masses.

As a further check that the magnitude gap provides information that is 
independent from richness, we have supplemented the comparison of 
the Mr19 groups to the MC groups by comparing the $\monetwo$ distribution 
of two subsamples of the Mr19 groups: 
one with $75$ $ \mathrm{km/s} < \sigma_{v} < 150 $ $\mathrm{km/s},$ 
and the other with $200 $ $\mathrm{km/s} < \sigma_{v} < 400 $ $\mathrm{km/s},$ 
both with the same distribution of richnesses. We find that at fixed richness, 
the groups with larger velocity dispersion have a higher proportion of large 
$\monetwo$ values relative to the small $\sigma_v$ groups such that the mean 
$\monetwo$ values of the two distributions are offset by $\sim 2.5\sigma.$ 
The sense of this trend is in keeping with the results presented in Fig.~\ref{fig:upshot}. 
This result explicitly demonstrates that $P(\monetwo|\sigma_{v},N)\ne P(\monetwo|N),$ 
directly implying that $P(\sigma_{v}|\monetwo,N)\ne P(\sigma_{v}|N),$ i.e., 
that the magnitude gap is informative about group/cluster mass even when the richness is known.

\subsection{Systematics Associated with Velocity Dispersion as Mass Proxy}
\label{sub:sigmavnoise}

We conclude this section by addressing several additional possible 
systematic errors in our data analysis, each related to our use of 
$\sigma_v$ as a mass proxy.  We first note that the design of the Monte Carlo comparison illustrated in the bottom right panel of Fig.~\ref{fig:upshot} addresses many of the concerns detailed below. Because each MC group inherits both the velocity dispersion and richness of its counterpart group in the data, a spurious separation between the observed large- and small-gap groups in Fig.~\ref{fig:upshot} due to $\sigma_v-$related systematics would be inherited by the MC groups. Thus in Fig.~\ref{fig:upshot}, the stronger separation in the data than in the Monte Carlos is unlikely to be due systematic errors related to noisy velocity dispersion measurements. 

We note, however, that our Monte Carlo sample is unrealistic in the sense that it makes no distinction between central galaxies and satellite galaxies, in discord with well-tested models based on the Conditional Luminosity Function formalism \citep[e.g.,][]{yang_etal07}. Moreover, while this Monte Carlo sample captures the basic features of the gap distribution remarkably well, it has been shown to fail to represent the magnitude gap distribution of our group sample in quantitative detail \citep{hearin_etal12a}. To address this shortcoming, we have performed a number of additional tests of possible $\sigma_v-$related systematics, which we now address in turn.

First, the two samples plotted 
in the bottom right panel of Fig.~\ref{fig:upshot} have different 
richness distributions, and so their velocity dispersions $\sigma_v$ 
are not determined with equivalent accuracy. 
In particular, as the error 
in $\sigma_v$ depends upon $N$, this may induce systematic 
differences between the measurements of the velocity dispersion of the 
groups in the two samples, potentially producing a spurious difference 
in the $\sigma_v-N$ relation exhibited by small-gap and large-gap systems.
To estimate the significance of this effect we have conducted the 
following exercise. For every group in each sample, we randomly select 
four members and use {\em only} these members to compute the velocity dispersion.  
We refer to the dispersion thus computed as the {\em reduced velocity dispersion}, 
$\sigma_v^{\mathrm{red}}$. For the small-gap system we find that the slope of the 
$\sigma_v-N$ relation is slightly steeper than the 
$\sigmavred-N$ relation (as expected, since the $\sigmavred$ measurements are 
noisier than $\sigma_v$ estimates), but the trend with gap persists at similar 
levels.  Our conclusion remains unchanged: 
at fixed $\sigmavred$, groups with a large magnitude gap 
have fewer members than groups with small values of $\monetwo.$ 

Second, r-band absolute magnitudes $\mr$, of the galaxies in our sample
have been estimated by using each galaxy's redshift. This introduces a
correlation between $\sigma_v$ and $\monetwo:$ groups with larger
velocity dispersions have a greater uncertainty in $\mr,$ which will
tend to enhance the small- and large-gap tails of the magnitude gap 
distributions in groups with larger $\sigma_v$ relative to groups with
smaller $\sigma_v.$ To test the significance of this effect, we have
recomputed $\mr$ for all the galaxies in our sample using 
the redshift of the group centroid and analyzed the 
galaxy sample with adjusted absolute r-band magnitudes. We find that
the change to all of our results is negligible, demonstrating that
this is not an important source of systematic error in our analysis.

Third, in calculating the velocity dispersion $\sigma_v$ of each group
we have made no distinction between central and satellite galaxies,
even though it is expected that $\sigma_v$ of centrals relative to the
group is much lower than that of the satellites. This choice
introduces an unphysical correlation between velocity dispersion and
richness since the contribution of the central galaxy to $\sigma_v$
decreases as the number of group members increases. To test for the
significance of this effect, we repeated our analysis when only
computing $\sigma_v$ by using the satellite members of each group. Of
course, we cannot unambiguously identify the galaxy closest the
potential minimum in each group, so we performed this test by assuming
that the object with the largest r-band luminosity is the central
galaxy. The analysis excluding the brightest galaxies within 
each group (as a proxy for the central galaxy) from the velocity dispersion 
estimate yielded results that differ only insignificantly from 
the primary analysis shown above. We conclude that this is 
not a significant source of systematic error in our results.


Fourth, there remains the possibility than when binning on true mass, small-gap and large-gap groups could in fact have the same mean richness (that is, it could be that the true signal we are trying to measure is actually zero), but that the map between true mass and $\sigma_v$ is different for large-gap and small-gap groups, inducing a spurious separation in $N-\sigma_v$ relation. To test for the possible influence of a gap-dependent map between mass and $\sigma_v,$ we use the mock catalogs studied in \S~\ref{section:predictions}. To remind the reader, the groups in this exercise are the host halos in the simulation, whose true mass is known. We begin with a sample of groups with a mass that is within $0.1$dex of some $M_{\mathrm{group}},$ and divide this fixed-mass sample into two subsamples: groups with small ($\monetwo<0.2$) and large ($\monetwo>1.5$) gaps. We find that for all group masses in the range $10^{13}\msun<M_{\mathrm{group}}<10^{14}\msun,$ the large- and small-gap subsamples always have commensurable $\sigma_v-$distributions (verified by a KS-test with p-values $>0.2$). This test confirms that the map between mass and $\sigma_v$ is not itself gap-dependent, so that using $\sigma_v$ as a mass proxy and selecting $\monetwo-$binned subsamples should not bias our mass-richness measurements.

Finally, we point out that a number of the potential systematics we have considered can be addressed if the data are binned on richness $N$ rather than $\sigma_v.$ For example, binning on richness eliminates the need to account for the effective $\sigma_v$ cut induced by our richness cut (see previous paragraph). Additionally, the purpose of the Monte Carlo comparison was to account for the statistical correlation between $\monetwo$ and richness; binning on $N$ eliminates the need for this comparison since, in this case, the $\monetwo-$selected group samples are being compared at fixed richness.

\begin{figure}
\centering
\includegraphics[width=8.0cm]{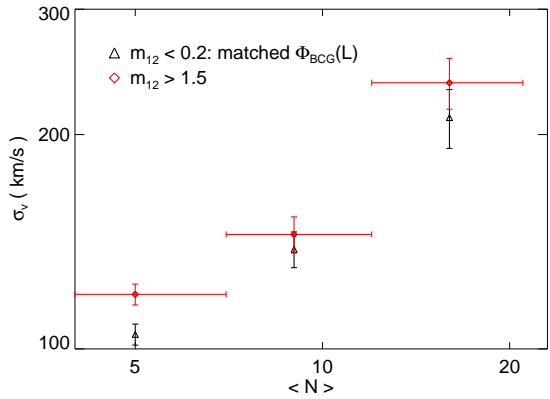}
\caption{
Alternative demonstration of the trend of the mass-richness relation with magnitude gap $\monetwo.$ This figure is analogous to the upper left panel of Fig.~\ref{fig:upshot}, except here we bin our groups on richness $N$ and measure the mean velocity dispersion $\langle\sigma_{v}\rangle$ in each richness bin. We plot results for large-gap groups with red diamonds, and small-gap groups with a matched $L_\bcg$ distribution with black triangles. Horizontal error bars indicate the boundaries of the richness bins, vertical error bars the error on the mean $\sigma_v$ in each richness bin. Large-gap groups exhibit larger velocity dispersions than small-gap groups at fixed richness, in keeping with the conclusions drawn from Fig.~\ref{fig:upshot}.
}
\label{fig:upshot2}
\end{figure}

We present the trend of the $N-\sigma_v$ scaling relation with magnitude gap $\monetwo$ in Figure \ref{fig:upshot2}. Within each richness bin, small-gap groups have a richer $N-$distribution than large-gap groups. To account for this effect, within each richness bin we randomly select subsamples of the small-gap groups with {\em both} a richness-distribution and a $L_{\bcg}$ distribution that matches that of the large-gap groups in that richness bin. We plot results for large-gap groups with red diamonds, and small-gap groups with a matched $N-$ and $L_\bcg-$distributions with black triangles. We have binned evenly in $\log N$ over the range $3\le N\le21,$ where the upper bound on $N$  is set by the largest richness exhibited by our $\monetwo\geq1.5$ group sample. Horizontal error bars in Fig.~\ref{fig:upshot2} (omitted from the plot of small-gap groups for visual clarity) indicate the boundaries of the richness bins, vertical error bars the error on the mean $\sigma_v.$ Adding in quadrature the errors on $\sigma_v$ from the two group samples, we find that the difference between the $\sigma_v$ distribution exhibited by large- and small-gap groups is $3.0\sigma$ discrepant from zero, with large-gap groups having larger velocity dispersions than small-gap groups at fixed richness. This provides an alternative demonstration that, at fixed richness, large-gap groups are more massive than small-gap groups, in keeping with our findings in which we bin instead on $\sigma_v.$ As discussed in \ref{section:conclusions}, a precise quantification of the improvement to cluster mass estimation provided by utilizing $\monetwo$ information will require the exploration of this technique in a photometric sample, which we leave as a task for future work.

\section{Summary \& Conclusions}
\label{section:conclusions}

We have studied the improvement to group and cluster mass determination that may 
be reaped by exploiting the magnitude gap, $\monetwo$, between the two brightest group and cluster members 
in addition to group richness, $N$.  
After fitting the mass-richness relation in our mock sample of 
simulated groups and clusters with a power law, we find a significant 
correlation between the magnitude gap and group mass residual, 
$\resid \propto 0.18\,\monetwo.$  The strength of this correlation is 
significant when compared to the scatter about our 
power law fit, $\sigma\left(\resid\right)=0.33.$ 

We see a similar trend in a volume-limited spectroscopic sample of galaxy groups 
observed in the SDSS. For group samples with different magnitude gaps, we find that 
large-gap groups have fewer members than small-gap groups at fixed $\sigma_{v}$. 
\citet{reyes_etal08} recently suggested that using the luminosity of the BCG could 
aid in reducing scatter in cluster mass determinations.  Similarly, we found that BCG luminosity 
is correlated with richness at fixed $\sigma_v$ in our SDSS groups.  
By constructing appropriate random samples from the SDSS data, we were able to 
conclude that $\monetwo$ contains information about 
group mass that is not contained in either richness or $L_{\bcg}.$ Our results are 
supported by conclusions drawn in \citet{ramella_etal07}, who find a correlation between 
magnitude gap and substructure abundance in the WINGS survey \citep{fasano_etal06}.

Our findings are closely related to a recent study by
\citet{paranjape_sheth11}, who employ a one-point statistical test to show that the abundance of galaxy
groups in Mr19\footnote{Note that the effective volume of the Mr19
galaxy group sample studied in \citet{paranjape_sheth11}, which was
based on SDSS Data Release 3, is less than half the effective volume
of our Data Release 7-based Mr19 sample.} as a function of  magnitude
gap is consistent with the distribution resulting from a set of random
draws from a global luminosity function, implying that group mass is
only related to $\monetwo$ through mutual covariance with
richness. Note, however, that \citet{paranjape_sheth11} also use a
marked correlation function analysis to show that this conclusion
cannot be entirely correct, and so it may not be surprising that our
findings are  in tension with their results that are based on one-point statistics.

Additionally, our results are in conflict with recent results claiming that the
distribution of magnitude gaps is determined purely by 
richness. \citet{proctor_etal11} claim that the {\em fossil
fraction}, defined as the fraction of groups with $\monetwo>2,$ is
purely a reflection of the abundance of low-richness systems. In
demonstrating that groups with different magnitude gaps exhibit
different relationships between $\sigma_{v}$ and richness we have
established that $P(\monetwo|\sigma_{v},N)\ne P(\monetwo|N),$
explicitly showing that the magnitude gap is not strictly determined
by richness.

We have demonstrated that, in principle, magnitude gap can inform cluster mass 
determination both in simple mock catalogs constructed from N-body simulations and 
in spectroscopic SDSS data; however, we leave the determination of the extent to which these relations may aid forthcoming 
cluster cosmology efforts as a subject of future work.  
Existing cluster cosmology samples differ from the group catalogs we have studied in several ways. 
For example, the {\tt maxBCG} clusters \citep{koester_etal07} are selected using photometric 
(rather than spectroscopic) data with an independent algorithm that uses color and i-band 
luminosity criteria. Additionally, the maxBCG clusters extend to higher luminosities and larger richnesses 
than our groups.  Forthcoming cluster cosmology efforts with data from imaging surveys 
like the Dark Energy Survey (DES) will likewise identify clusters using photometric data 
and probe larger richnesses.  Moreover, the richnesses of maxBCG clusters are defined 
according to a more complex optimization procedure than the 
richnesses of our groups \citep{koester_etal07,rozo_etal09}.  
We have analyzed the \citet{berlind_etal06} clusters because the cluster membership 
assignments for the {\tt maxBCG} clusters are not publicly available.  We do not 
anticipate the correlations that we point out here to be particularly strong functions 
of redshift or richness, but this will need to be tested more extensively both in 
mock catalogs and forthcoming data.

There are at least two distinct ways in which $\monetwo$ may be exploited 
to tighten the relationship between group/cluster mass and richness.  
First, the magnitude gap could be treated on an equal footing with richness, 
so that rather than calibrating the mass as a one-dimensional function of richness 
one could instead treat the mass as a function of $N$ and $\monetwo$ simultaneously. 
Simulations coupled with detailed studies of extant and near-future data could provide 
parameterized forms for the $\monetwo-M$ relation with reasonable priors as they do now 
for the $N-M$ relation.  We studied the potential benefit of this approach in our mock 
group sample by comparing the difference in the scatter about the residual mass estimation 
between a one-dimensional linear regression on richness and a two-dimensional linear 
regression on richness and magnitude gap. We find that the scatter in the residuals $\resid$ 
improves by $\sim 15-20\%$ when using a fit for mass as a function of N and $\monetwo$ instead of N alone.
While this improvement may seem modest, it comes at no additional observational cost, because 
the magnitude gap will always be available in the same data set used to measure the richness.

A second approach is suggested by the observation that the relationship 
between $\resid$ and $\monetwo$ in our mocks appears to be nonlinear, with the 
systems with the very largest gaps appearing to be outliers in the mass-richness relation 
(see Fig.~\ref{fig:residual}). The highest-gap systems have inordinately large masses 
at fixed richness.  One may imagine utilizing gap information to identify significant 
outliers in the mass-richness relation.  It may be possible to impose a cut on 
$\monetwo$ and restrict consideration to groups and clusters with a modest magnitude gap 
($\monetwo\lesssim1.5$) to calibrate the mass-richness relation.  This may be particularly 
helpful in photometrically-identified groups because interloper contamination will be 
more significant in the absence of spectroscopic redshifts, but interlopers can only 
{\em reduce} the magnitude gap, so large-gap systems will remain mass-richness outliers.  
Of course, the effect of any such cut on cluster abundance must be calibrated and 
accounted for, and this must be a subject for further work. In a forthcoming companion paper 
(Hearin et al.~2012, in prep), we study in detail the global abundance of groups and 
clusters as a function of $\monetwo$ over a wide range of masses, providing precisely 
the information that would be required to carry out this second approach. 

As we were preparing to submit this paper for publication, we became
aware of another paper \citep{wu_etal12}  that studies some of the
same material that we have addressed. In particular, they focused on
the host halo with the most extreme difference between the $\vmax$
value of the host and its largest subhalo, the N-body simulation
analog of a cluster with a very large magnitude gap. They found that
this halo also appeared to be an outlier in many of the host halo
properties they studied, including the number of subhalos contained by
the host. This finding appears to be in keeping with the second
approach described above to using $\monetwo$ in the calibration of the
mass-richness relation.

However the calibration is conducted, our results suggest that the magnitude gap 
$\monetwo$ contains information about cluster mass that is independent from both 
richness and $L_{\mathrm{BCG}},$ the luminosity of the brightest cluster member. 
Exploiting this additional information to improve existing optical tracers of 
cluster mass may improve the constraining power of optically-identified galaxy 
clusters on cosmology.

\section*{Acknowledgments}

We are grateful to Nick Battaglia, Ted Bunn, Surhud More, Eduardo Rozo, 
and Anja Weyant for helpful discussions, and 
to Peter Behroozi and Risa Wechsler for making  available the {\tt ROCKSTAR} halo catalogs 
for the {\tt Bolshoi} simulation.  APH and ARZ are supported in part by 
the National Science Foundation (NSF) through grants NSF AST 0806367 and NSF AST 1108802.  
APH has been supported in part by the the Pittsburgh Particle physics, Astrophysics, and Cosmology 
Center (PITT PACC) at the University of Pittsburgh.  AAB is supported by the Alfred P. Sloan Foundation, 
as well as the National Science Foundation through grant NSF-AST 1109789. JAN is supported by the 
United States Department of Energy Early Career program via grant DE-SC0003960
and NSF AST grant 08-06732.

\bibliography{gaprich}

\end{document}